\documentclass[runningheads]{llncs}
\usepackage{graphicx}
\usepackage{hyperref}
\usepackage[T1]{fontenc}
\usepackage{multirow}
\usepackage{booktabs}
\usepackage{amsmath} 

\begin{document}
\title{Longformer for MS MARCO Document Re-ranking Task}
%
%
\author{Ivan Sekuli\'c\inst{1} \and
Amir Soleimani\inst{2} \and
Mohammad Aliannejadi\inst{2} \and
Fabio Crestani\inst{1}}
\authorrunning{I. Sekuli\'c et al.}
%
\institute{Universit\`a della Svizzera italiana \email{name.surname@usi.ch}\and
University of Amsterdam \email{n.surname@uva.nl}}
\maketitle              
\begin{abstract}
Two step document ranking, where the initial retrieval is done by a classical information retrieval method, followed by neural re-ranking model, is the new standard.
The best performance is achieved by using transformer-based models as re-rankers, e.g., BERT.
We employ Longformer, a BERT-like model for long documents, on the MS MARCO document re-ranking task.
The complete code used for training the model can be found on: 
\url{https://github.com/isekulic/longformer-marco}

\keywords{Document ranking \and Longformer \and Neural ranking.}
\end{abstract}
\section{Introduction}
Document ranking is central problems in information retrieval (IR).
Given a query, the task is to rank the documents of some collection so that the most relevant ones appear on top of the list.
Recently, neural ranking models have shown superior performance compare to the traditional IR methods.
Given the much higher computational need of neural models, the two step retrieval process is widely adapted.
First, a traditional IR method, like BM25 or query likelihood, retrieves top $k$ documents from a given collection.
Then, a computationaly expensive neural model re-ranks the top $k$ documents from the initial retrieval step.

A number of neural models for ranking has been proposed in recent years. Some of them are: DRMM \cite{guo}, KNRM \cite{xiong}, Co-PACRR \cite{hui}, DUET \cite{mitra}, and Conformer-Kernel with QTI \cite{conformer}.
With a combination of new generation of neural models, namely the transformer architecture, and large-scale datasets, neural rankers arose as superior to traditional methods, which was not possible before \cite{lin}.
Most notable model from the transformers family is probably BERT \cite{bert}, which has been very successfully applied to passage ranking \cite{marco_bert}, outperforming state-of-the-art by a large margin. 

The largest dataset for the document ranking task is MS MARCO (Microsoft Machine Reading Comprehension).
Transformer architecture, namely BERT, has already proven effective on the MS MARCO passage ranking task.
However, the documents are much longer than the passages, making the task of document ranking more challenging.

To address the issue of increased length of the documents, we employ Longformer \cite{longformer} -- a BERT-like model for long documents.
Longformer has an adjusted attention mechanism that combines local, BERT-like windowed attention, with a global attention, allowing the model to attend over much longer sequences than standard self-attention.
Compared to BERT that typicaly processes up to $512$ tokens at a time, Longformer is pre-trained on documents with length of $4096$ tokens.
We reach MRR@100 of $0.329$ and $0.305$ on the official dev and the test sets, respectively.

\section{Dataset}
We train and evaluate Longformer on MS MARCO document ranking dataset.
It consists of more than 370k queries, 3.2 million documents, and the corresponding relevance labels for query-document pairs.
Relevance labels are transferred from the MS MARCO passage ranking task, by mapping a positive passage to a document containing the passage, for each query.
This is done under an assumption that the document that contains a relevant passage is a relevant document.
Additional information about the dataset is present in Table \ref{tbl:marco}.

\begin{table}[]
\centering
\begin{tabular}{llr}
\toprule
\# documents                &       & 3.2M    \\
\midrule
\multirow{3}{*}{\# queries} & train & 367,013 \\
                            & dev   & 5,193   \\
                            & test  & 5,793  \\
\bottomrule \\
\end{tabular}
\caption{Number of documents in MS MARCO corpus and the number of queries in train, dev, and test set.}
\label{tbl:marco}
\end{table}

The document ranking task features two tasks:
\begin{description}
    \item[Document Re-Ranking] Given a candidate top 100 documents for each query, as retrieved by BM25, re-rank the documents by relevance.
    \item[Document Full Ranking] Given a corpus of 3.2m documents generate a candidate top 100 documents for each query, sorted by relevance.
\end{description}
We participate in the document re-ranking task, where we use Longformer to assign relevance to the 100 documents retrieved by BM25, which are provided by the organizers.

\section{Experiments}
We train Longformer \cite{longformer} to estimate relevance of a query-document pair. 
The training setting is formulated as in \cite{marco_bert}. 
We feed the query as sentence A and the document as sentence B to the tokenizer, which yields the following input to the Longformer\footnote{Tokens <s> and </s> are equivalent to the [CLS] and [SEP] tokens in BERT tokenizer, respectively.}:
\begingroup
\setlength{\thickmuskip}{0mu}
\begin{equation*}
    <s>\  query\  </s>\  document\  </s>
\end{equation*}
\endgroup

We truncate the document such that the sequence of the concatenated query, document, and the seperator tokens does not exceed $4096$ tokens. 
After passing the sequence through the Longformer model, the\begingroup
\setlength{\thickmuskip}{0mu}
<s>
\endgroup vector is given as input to a classifier head, consisting of two linear layers with dropout and a non-linear function. 
The classifier outputs a two-dimensional vector, where the first dimension indicates probability of a document not being relevant to the query, while the second indicates relevance probability.
For a given query, we rank the candidate documents based on the relevance probability, which is computed independently for each document.

We fine-tune the pre-trained Longformer with a cross-entropy loss, using the following hyperparameters: batch size of $128$, Adam optimizer with the initial learning rate of $3\times10^{-5}$, a linear scheduler with warmup for $2500$ steps, trained for 150k iterations. 
Further hyperparameter tuning might yield better results. 
For training, we also reduce the positive to negative document ratio to 1:10, from the given 1:100 (as each query is given top 100 documents extracted by BM25 by the organizers). 
We use PytorchLightning \cite{pl} for our training setting implementation and HuggingFace's Transformers package \cite{hug} for Longformer implementation.

\section{Results}
The results on the MS MARCO document re-ranking task on the dev and the test set are presented in Table \ref{tab:results}.
The official metric is mean reciprocal rank (MRR@100).
Other submissions and approaches can be found on the official leaderboard\footnote{https://microsoft.github.io/msmarco/\#docranking}.

\begin{table}[]
\centering
\begin{tabular}{lrr}
\toprule
                                  & dev         & test  \\
\midrule 
Indri Query Likelihood            &             & 0.192        \\
Conformer-kernel with QTI (NDRM3) &             & 0.293        \\
Conformer-kernel with QTI (NDRM1) &             & 0.307        \\
Longformer                        & 0.336       & 0.305       \\
\bottomrule \\
\end{tabular}
\caption{MRR@100 of the Longformer and the official baselines provided by the organizers. \cite{conformer}}
\label{tab:results}
\end{table}

%
%
%
%

\end{document}